\documentclass[aip,sd,amsmath,amssymb,reprint]{revtex4-1}
\usepackage[pdftex]{graphicx} 
\usepackage{dcolumn}
\usepackage{bm}
\graphicspath{{../pdf/}{../jpeg/}{./image/}}    
   \DeclareGraphicsExtensions{.pdf,.jpeg,.png,.jpg}

\begin{document}

\preprint{AIP/123-QED}

\title{Metallic multilayers for X-band Bragg reflector applications}

\author{A.P. Mihai}
\email{a.mihai@ic.ac.uk}

\author{M. Adabi}
\author{W. Liu}

\author{H.Hill}
\author{N. Klein}
\author{P. K. Petrov}
 
\affiliation{Department of Materials, Imperial College, London SW7 2BP, United Kingdom}%

\date{\today}

\begin{abstract}
We present a structural and high frequency (8.72GHz) electrical characterization of sputter deposited Ti/W, Ti/Ru and Mo/Ti metallic multilayers for potential application as acoustic Bragg reflectors. We prove that all metallic multilayers comprised of different acoustic impedance metals such as Ti, W, Mo are promising candidates for Bragg reflector/bottom electrode in full X-band thin film acoustic resonators. Values for high frequency resistivity of the order of $10^{-8}$ ohm.m are measured by use of a contact-free/non-invasive sheet resistance method.
\end{abstract}


\maketitle 



Thin film varactors and acoustic resonators with electric-field tunable permittivity of ferroelectric perovskite oxides (e.g., BaxSr1-xTiO3)
have been extensively studied for applications in microwave technology.\cite{Noe2007, Zhu2009} Such devices are typically fabricated in an interdigital capacitor or metal-insulator-metal (MIM) geometry. In MIM microvawe devices, "`metal"'denotes the conducting bottom and top electrodes of the heterostructure. The choice of bottom electrode material is critical to achieve a high quality factor (both acoustic and electric) and tunability of the MIM microwave, as it affects the microstructure of the dielectric (I) layer\cite{Vor2011}, and properties of the M-I interface.

In the special case of thin film bulk acoustic resonators (TFBAR), two build geometries are generally used. One relies on a free standing membrane geometry\cite{Noe2007} and the second one relies on pre-deposition of a multilayer consisting of alternating low and high acoustic impedance materials\cite{Yan2012}.  The membrane geometry provides a good acoustic quality factor for the resonances but presents a fabrication challenge as complicated etching techniques need to be used in order to construct the cavities. These multilayers provide a solution, but lower quality factors are measured, as structural quality is very important with regard to possible losses mechanisms.

These multilayers are called acoustic Bragg reflectors, and they provide acoustic wave insulation of the piezoelectric active layer from the substrate. Usually acoustic Bragg reflectors consisting of an oxide (SiO2,TaO2) and a high acoustic impedance metal (W) are being used by other groups. 

Generally metals such as Pt and Au have been used as bottom electrodes due to their low resistivity (low losses). Also, in general the role of the Bragg reflector is to just provide the acoustic insulation and not to be used as a conductive bottom electrode.

There have been reports of all metallic Bragg reflectors \cite{wei2008,Che2010}, but no high frequency results on their resistivity have been reported, as they haven’t generally been used as bottom electrodes.

In this paper we present structural and electrical transport results from all metallic multilayers based on Ti combined with Ru, Mo and W made by sputtering at room temperature. We focus on a non intrusive way of measuring microwave electrical behaviour of these multilayers, and conclude that these materials could provide a low cost, rapid solution of integrating both acoustic insulation and electrical conduction into one heterostructure.


A series of metallic multilayers of different acoustic impedance have been grown on thermally oxidized Si/SiO2 substrates by use of a sputtering chamber. The base vacuum of the chamber was of the order of $10^{-7}$mbar with a $10^{-3}$mbar growth pressure. The substrates were cleaned and dried by use of a standard method. Sputtering was done in a pure Argon atmosphere without any substrate heating. The different materials were sputtered from 3 inch diameter $>99.95$ pure targets (Pi-kem).

Layer crystallinity and orientation of as grown samples was analyzed by XRD. Surface roughness was measured by use of AFM (Innova..etc), then slice views of each sample were obtained by use of an SEM (Leo Gemini 1525 FEGSEM). DC resistivity measurements were performed at room temperature by a 4-probe technique (Keithley 2400). 

High frequency microwave measurements at room temperature  were performed by use of a Rhode Schwartz Vector Network Analyzer (VNA) setup at a frequency of 8.72GHz. In this technique a high dielectric constant (29) and a very low loss tangent (less than 0.0001 at microwave frequencies) cylindrical puck made of Barium Zinc Titanate (BZT) ceramic is used. It is enclosed inside a copper cavity, where it sits on an adjustable quartz holder in the center of the cavity. The top surface of the ceramic resonator is separated from the upper cavity wall by 1mm.  The resonant field is coupled to the structure by means of two coupling loop antennas inside the cavity which are fed by a state of art VNA. The transmitted (S21) and reflected (S11) power are measured as a function of frequency. The $TE 01\delta$ mode is chosen for the measurements as it has azimuthal  symmetry of the electromagnetic field associated circular current excited in the samples through the evanescent field that leaks through the designated 5mm aperture, over which the sample is staged.

Three combinations of metals have been chosen. Ti has a low acoustic impedance and is commonly used as an adhesion layer in thin film deposition. Ru, Mo and W on the other hand have high acoustic impedance (refer to table 1), and are also high temperature structurally resistive. The necessity for these metallic layers to be high temperature and oxidation resistive stems form the necessity of using high temperature deposition techniques for subsequent piezoelectric layer deposition. Vorobiev et al.\cite{Vor2011} have shown that the thermal expansion coefficient of the materials that are chosen to provide the Bragg acoustic insulation is important, and in consequence we have chosen metals with similar thermal expansion coefficients. 

Thus, three material combinations were chosen and deposited the corresponding thicknesses in order to obtain an acoustic insulation from the substrate that would be centered at around 7GHz . The combinations are Si(SiO2)/(Ti/Ru)x3, Si(SiO2)/(Ti/W)x3, Si(SiO2)/(Mo/Ti)x3. Values for acoustic impedance, thermal expansion coefficient and corresponding layer thicknesses are found in table 1.Previous work has proven that, by increasing the number of repetitions of the two different acoustic impedance layers, an increase in acoustic insulation and quality factor of the resonances is observed \cite{wei2008}. We have chosen to settle for 3 repetitions as the benefit of increasing this number is only marginal\cite{wei2008} (loss of bandwith with higher number of repetitions).

\begin{table}[!h]
	\centering
		\begin{tabular}{|c|p{1.5cm}|p{2cm}|c|}
			\hline Material & Acoustic Imp. (MRayl) & Thermal exp. coeff.  ($10^{-6}m/mK$) & Thickness, $\lambda/4 (nm)$ \\
			 \hline Ti & 27 & 8.6 & 182  \\
			        W & 102 & 4.5 & 165  \\
							Ru & 67 & 9.1 & 217  \\
							Mo & 63 & 4.8 & 181\\
							\hline

		\end{tabular}
	\caption{Acoustic impedance, thermal expansion coefficient and nominal layer thicknesses calculated for a 7GHz centered acoustic Bragg reflection band.}
	\label{tab1}
\end{table}


All samples are polycrystalline with a variable degree of crystalinity (xrd results not shown), as expected from a sputtering growth method (at room temperature). One can easily see the corresponding pics depending on the combination of materials used. Ti is present in all samples with (100) and (002) crystalline orientations. Ru displays a two phase structure with (110) and (110) pics, while Mo only shows a (101) pic. 


SEM slice views of the three types of multilayers are presented in figure 1. From the SEM slice views one can conclude that the layers are continuous with a polycrystalline morphology. Interfaces are smooth and continuous and no mixing of different metals has been observed. The measured surface roughness are 11nm/9nm/8.5nm for Ru/W/Mo containing samples respectively. Given that the total thickness of these multilayers is around 1/mu m, a surface roughness of about 10nm is indeed a positive result. Surface roughness of the metallic multilayers is important for subsequent processing as the eventual piezoelectric layer must have minimal roughness in order to minimize losses. Attenuation coefficient for transverse roughness much smaller than the wavelength is directly proportional to the layer roughness squared\cite{Vor2011}, hence for an active layer thickness of about $100-150nm$ a roughness of under 10nm is desired.

\begin{figure*}[!ht]
\includegraphics[trim=0 0 0 0,clip, width=16cm, height=9cm]{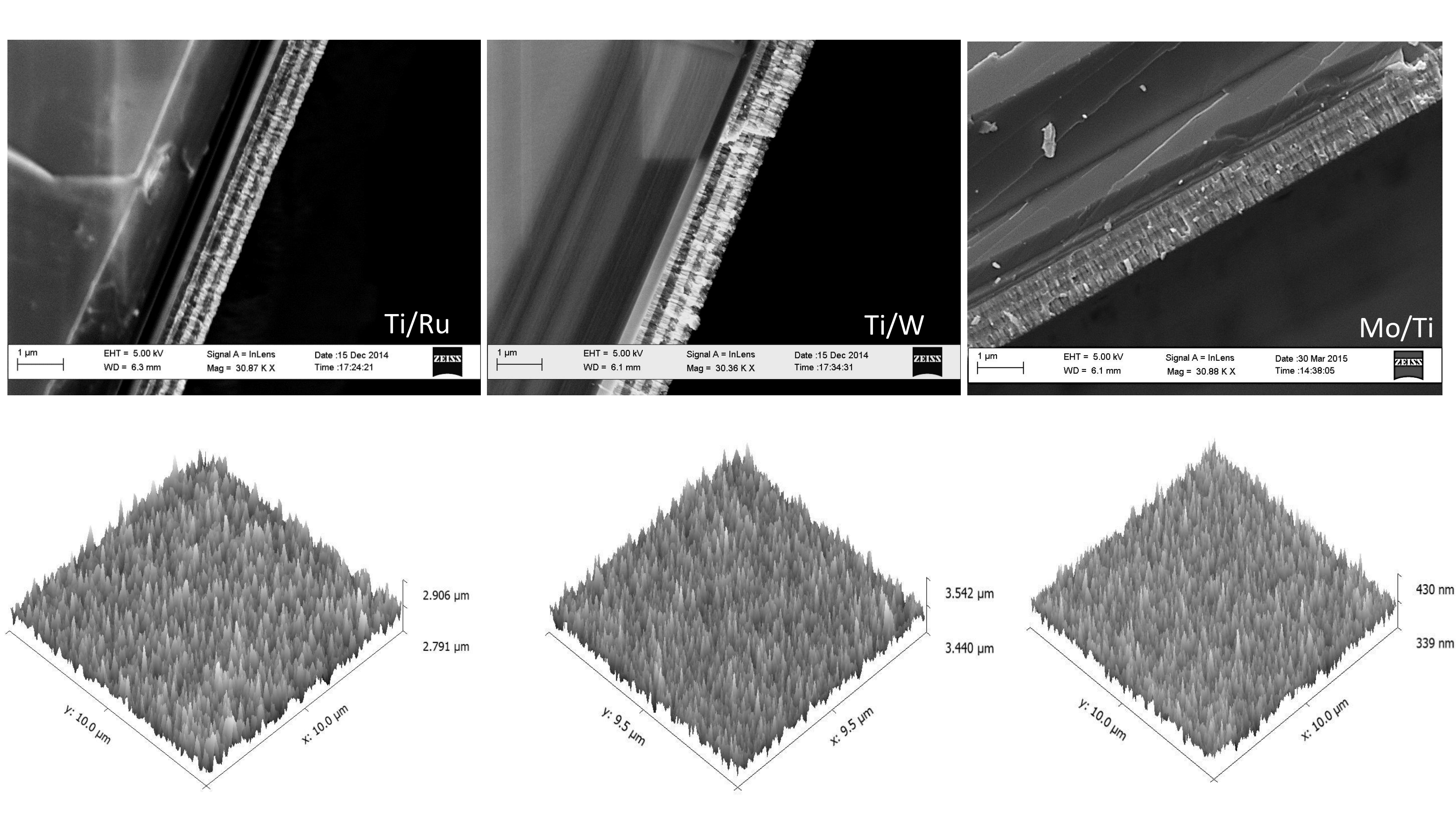}
\caption{\label{fig1} SEM and AFM scans of the three types of samples.}
\end{figure*}

Room temperature DC resistivity/conductivity measurements have been performed by a classic 4-probe method and results are summarized in table 2. On three samples we have also grown a 50nm Pt layer as  further protection against oxidation. No significant change was noticed in the measured resistivity/conductivity by the addition of a Pt layer. In terms of DC resistivity measured at room temperature, all combinations of metals chosen display values of the order of $10^{-6} ohm.m$, with no real material dependence, which shows that electronic transport is governed mostly by scattering at interfaces, given the polycrystalline nature of our films.

\begin{table}[!ht]
	\centering
		\begin{tabular}{|c|p{1.5cm}|p{2cm}|c|}
			\hline Material & Resistance (ohm) & Resistivity (ohm.m) & $\sigma$ (S/m) \\
			 \hline (Ti/W)x3 & 1.36 & 7.4E-06 & 1.4E+05\\
			        (Ti/Ru)x3 & 0.84 & 4.6E-06 & 2.2E+05 \\
							(Mo/Ti)x3 & 1.30 & 7.1E-06 & 1.4E+05\\
							(Ti/W)x3+50nm Pt & 1.30 & 7.4E-06 & 1.4E+05\\
							(Ti/Ru)x3+50nm Pt & 0.73 & 4.1E-06 & 2.4E+05\\ 
							(Mo/Ti)x3+50nm Pt & 1.30 & 7.4E-06 & 1.3E+05\\
							\hline

		\end{tabular}
	\caption{Room temperature DC resistance and corresponding resistivity/conductivity for all three types of samples.}
	\label{tab1}
\end{table}

In the following we will focus on high frequency electronic transport behaviour of our thin film multilayers.

\begin{table*}[!ht]
	\centering
		\begin{tabular}{|p{3cm}|p{2cm}|p{2.2cm}|p{2cm}| p{2.5cm}|c|c|}
			\hline DC Resistance (ohm) & DC Sheet Resistance (ohm/sq) & DC Resistivity (ohm.m) & Skin depth (m) & Sheet Resistance at $8.72$ Ghz (ohm/sq)  & $\Delta Q^{-1}$ & $\alpha (\Omega^{-1}))$ \\
			\hline 5.50E-02 & 0.25 & 3.99E-08 & 1.08E-06 & 0.037 & 1.63E-05 & 6.05E-06 \\
			\hline
		\end{tabular}
	\caption{Calibration of measurement using 160nm of Sputtered gold on polished quartz substrate.}
	\label{tab1}
\end{table*}

A dielectric resonator method has been successfully employed for RF testing of high temperature superconducting films \cite{}[11-14], superconducting plates \cite{Maz97}, epitaxial thin films \cite{Kle92}, lossy liquids \cite{Sha2015,Bar2007}, ferroelectrics \cite{Kru2003}, as well as contactless sheet resistance measurements of graphene on low-loss substrates \cite{Sha2008}.

The resonant frequency of the setup with no samples is measured to be at around 8.7235 GHz with a Q-factor (electrical factor) of 9220. Once a conductive sample is put in place over the aperture, there will be a shift in frequency to lower values and a drop in Q-factor due to the induced current in the conductive film which results in perturbation of the field. For a given conductive material with a known AC sheet resistance, that has thickness of about an order of magnitude larger than its skin depth, sheet resistance at the operation frequency is related to the change in Q-factor of the resonator with and without the conducting sample via eq.1: 

\begin{equation}
\label{equation1}
\Delta(Q^{-1} )=  1/Q(Substrate) -1/Q(empty) =\alpha R_{s}
\end {equation}

The sheet resistance at a given frequency can be found from the skin depth of that isotropic material at that frequency:  

\begin{equation}
\label{equation2}
R_{s}=\frac{\delta}{\rho}=\frac{\sqrt{\frac{\rho}{\pi f \mu_{0}}}}{\rho}=\sqrt{\frac{1}{\pi f \mu_{0}\rho}}
\end {equation}

The coefficient $\alpha$ is found by sputter deposition of 160nm gold on a $1cm^{2}$ quartz substrate and using 4-point probe measurement to find its resistivity combined with the change in Q-factor investigated using the method explained earlier. 

CST Microwave Studio\cite{cst} software was used and change in electric and magnetic fields was simulated. This was performed for $1.2\mu m$ thin film with DC conductivity of 0.135MS/m deposited on high resistivity silicon substrate. As can be seen, introduction of the sample results in interaction between the evanescent field that leaks through the aperture and the film (see figure 2). This interaction results in change in Q-factor of the resonator. (top row left (E-field) right (H-field) of unloaded cavity, bottom row left (E-field) right (H-field) of loaded cavity with conductive stack of films on HRS). 

\begin{figure*}[!ht]
\includegraphics[trim=0 0 0 0,clip,width=10cm, height=6cm]{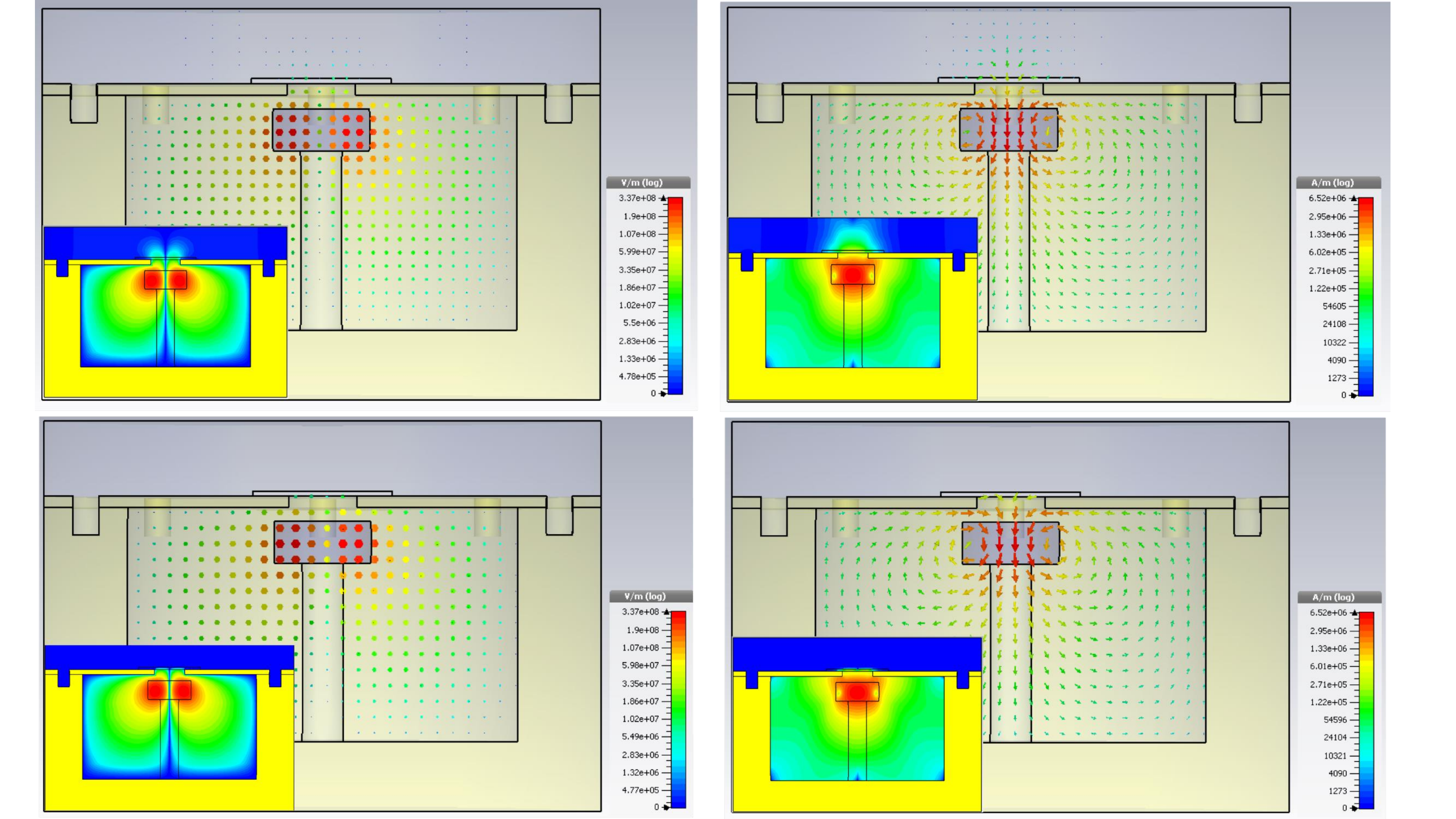}
\caption{\label{fig4} Simulated changes in electric and magnetic fields for different configurations. Top two unloaded cavity; bottom loaded cavity.}
\end{figure*}

Q-factor measurements were repeated using samples containing the stacks of thin films with total thickness of 1.2µm (see figure 3). As can be seen, samples fabricated under same conditions yield very close resistivity values indicative of the high accuracy of the method. Equation 1 along with the measured Q-factor for each sample were used and resistivity of each stack was found at the frequency of about 8.7GHz. Comparing the obtained AC values to those measured at DC, it is apparent that microwave resistivity values are lower by about two orders of magnitude. It is also interesting to note that the order of most to least conductive stacks is different in AC compared to DC. The least conductive stack at DC is found to be T3W3 whereas M3T3 has the highest resistivity at AC.  Generally the current distribution in AC is going through all the layers whereas DC will have inhomogeneous current distribution depending on the position of the probes and relative resistivity of each layer, whereas in AC the field penetrates all the layers and field is distributed homogeneously (as shown in figure 2).

\begin{figure}[!hb]
\includegraphics[trim=50 0 50 50,clip,width=9cm, height=5cm]{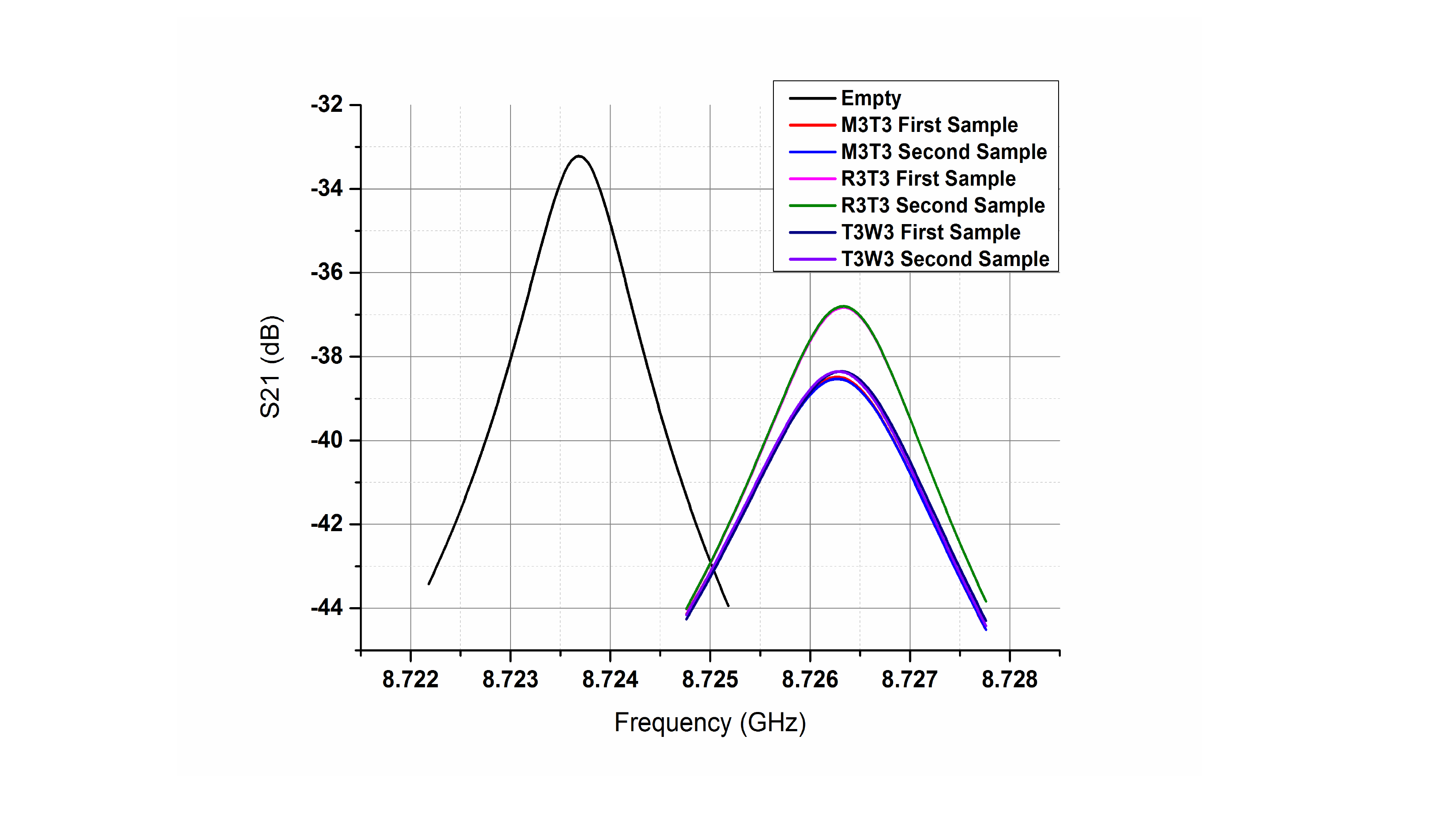}
\caption{\label{fig3} (color online) S21 frequency shift results for different multilayer combinations.}
\end{figure}

A change in the resistivity at high frequency is expected, but normally in metals the frequency values for which resistivity starts to drop are found in the near IR region \cite{tas2012}. Generally used microwave metals such as Au, Pt and Cu display plasma frequencies in the Terra-hertz region (and are commonly used as microwave electrodes) but it is generally accepted that "`bad"' metals and alloys can show a drop in resistivity in the microwave region \cite{tas2012}. Structural quality and roughness have been cited as reasons for increased high frequency resistivity at microwave frequency \cite{Run2003}. We have found no real dependence of the AC resistivity with regards to surface morphology, leading us to believe that the overall microstructure is not the determining factor in microwave electronic transport in the case of such metallic multilayers.

\begin{table}[!h]
	\centering
		\begin{tabular}{|p{1.5cm}|p{1.5cm}|p{1.7cm}|p{1.5cm}|p{1.5cm}|}
			\hline Sample & $\Delta Q^{-1}$  & Sheet Resistance at 8.72GHz (ohm/sq) & Resistivity at 8.72 Ghz (ohm.m) & DC Resistivity (ohm.m)\\
			\hline T3W3 (sample1) & 2.71E-04 & 0.0614 & 7.37E-08 & 7.4E-06 \\
			T3W3 (sample2) & 2.64E-04 & 0.0598 & 7.19E-08 & 7.4E-06 \\
			 T3R3 (sample1) & 2.02E-04 & 0.0459 & 5.50E-08 & 4.6E-06 \\
			 T3R3 (sample2) & 2.02E-04 & 0.0458 & 5.50E-08 & 4.6E-06 \\
			 M3T3 (sample1) & 2.81E-04 & 0.0638 & 7.66E-05 & 7.1E-06 \\
			 M3T3 (sample2) & 2.81E-04 & 0.0639 & 7.66E-08 & 7.1E-06 \\
			\hline
		\end{tabular}
	\caption{Resistivity calculations at Microwave frequency for all three material combinations.}
	\label{tab2}
\end{table}

A good alternative to metallic electrodes would be highly epitaxial electrode layers ($SrMoO_{3}$) with low defect density. They have been shown to be very good candidates also for microwave applications\cite{Rad2014}, but this requires high temperature MBE like deposition techniques which hinder their technological application. 


In this paper we have presented structural, DC and microwave electrical transport results of room temperature sputter deposited Ti/Ru, Mo/Ti and Ti/W metallic multilayers with potential application as metallic acoustic Bragg reflectors. From the high frequency measurements we can conclude that in terms of resistivity, a value of the order of $10^{-8}ohm.m$ is sufficient for use as a microwave devices electrode. No real dependance on surface roughness has been observed, even though a further improvement of the microstructure and roughness is desired.



%


\bibliography{Bragg}
\end{document}